\documentclass[aps,10pt,prd,twocolumn,preprintnumbers,amsmath,amssymb,nofootinbib,a4paper,showpacs]{revtex4-1}

\usepackage{amsfonts,amsthm}
\usepackage{bm,bbm}
\usepackage{graphicx}
\usepackage[T1]{fontenc}
\usepackage{color}
\usepackage[breaklinks]{hyperref}

\newcommand{\be}{\begin{equation}}
\newcommand{\ee}{\end{equation}}
\newcommand{\ba}{\begin{eqnarray}}
\newcommand{\ea}{\end{eqnarray}}
\def\bs{\begin{subequations}}
\def\es{\end{subequations}}

\def\de{\delta}
\def\g{\gamma}
\def\la{\lambda}

\def\Om{\Omega}
\def\om{\omega}

\def\t{\tau}  
\def\s{\sigma}
\def\vr{\varrho}

\def\cA{\mathcal{A}}

\def\cD{\mathcal{D}}

\def\cF{\mathcal{F}}

\def\cJ{\mathcal{J}}

\def\ds{d_{\rm S}}
\def\dh{d_{\rm H}}

\def\p{\partial}

\newcommand{\Eq}[1]{(\ref{#1})}
\def\com{\color{magenta}}
\def\cob{\color{blue}}
\newcommand{\book}[5]{\emph{#1}  (#2, #3, #4, #5)}
\newcommand{\books}[4]{\emph{#1} (#2, #3, #4)}

\newcommand{\arX}[1]{\href{http://arxiv.org/abs/#1}{{\ttfamily\com arXiv:#1}}}
\newcommand{\doin}[6]{\href{http://dx.doi.org/#1}{{\cob #2 #3 {\bf #4}, #5 (#6)}}}
\newcommand{\doinn}[5]{\href{http://dx.doi.org/#1}{{\cob #2 {\bf #3}, #4 (#5)}}}
\newcommand{\doij}[5]{\href{http://dx.doi.org/#1}{{\cob #2 #3 (#5) #4}}}

\newcommand{\ndoinn}[5]{\href{#1}{{\cob #2 {\bf #3}, #4 (#5)}}}
\newcommand{\tia}[1]{}

\def\rmd{d}

\begin{document}

\title{Relativistic particle in multiscale spacetimes}

\author{Gianluca Calcagni}
\email{calcagni@iem.cfmac.csic.es}
\affiliation{Instituto de Estructura de la Materia, CSIC, Serrano 121, 28006 Madrid, Spain}

\begin{abstract}
We study the action and the dynamics of a relativistic particle, uncharged or charged, in multiscale spacetimes. Invariance under reparametrizations and Poincar\'e symmetries uniquely determine the action and the line element to be the usual ones, without the weight factors typical of particle mechanics in these geometries. The resulting spacetime is multiscale only along spatial directions. This version of the system is also dictated by recovery of the nonrelativistic limit together with compatibility with Maxwell and electrodynamics field theory. Giving up all these requirements and allowing for a nontrivial weight factor in the time direction produces a modified line element and considerably complicates the dynamics in the case of a charged particle.
\end{abstract}

\date{June 23, 2013}



\preprint{\doin{10.1103/PhysRevD.88.065005}{Phys.\ Rev.}{D}{88}{065005}{2013} \hspace{10.5cm} \arX{1306.5965}}

\maketitle


\section{Introduction}

Recently, spacetimes whose geometry changes with the probed scale have been introduced (see \cite{frc1,frc2,frc6,frc7} and references therein). Dynamics is defined through an action of the form $S=\int\rmd\vr(x)\,\mathcal{L}$, where $\vr(x)$ is a Lebesgue--Stieltjes measure with anomalous scaling $\vr(\la x)\sim\la^{\dh}\vr(x)$. The anomalous exponent $\dh$ is the Hausdorff dimension of spacetime and its value is, in general, scale dependent and different from the integer topological dimension $D$. Assuming that $\rmd\vr(x)=d^Dx\,v(x)$ and that the weight $v$ is positive semidefinite and factorizable in the coordinates, $v(x)=\prod_{\mu=0}^{D-1}v_\mu(x^\mu)$, one can realize anomalous and multiscale geometries via multifractional measures, i.e., such that each $v_\mu$ is the sum of noninteger powers of the coordinates. The purpose of fractional models is twofold. On one hand, to understand renormalization properties of field theories living in nonconventional geometries \cite{frc2,frc9}. On the other hand, to better characterize effective spacetimes in quantum gravity by introducing an ``alternative toolbox'' of instruments borrowed from multifractal geometry \cite{frc1,frc2} and transport and probability theory \cite{frc7,frc4} (all fields where multifractional measures play an important role). Anomalous geometries arise most naturally in quantum gravity, either by construction (in the quest for theories with absent or tamed infinities in the ultraviolet) or from quantum effects, or both. Then, multiscale geometries and the tools of fractional calculus may serve as effective models describing certain regimes or features of other theories \cite{ACOS,fra7,CES}.

Presently, we will continue this trend of investigation by examining in detail the relativistic particle. The particle's worldline is described by a parametrization of the spacetime position coordinates $x^\mu(\t)$, which are the fundamental variables of the problem. As in standard Minkowski spacetime, imposing the action to be invariant under \emph{ordinary} Lorentz or Poincar\'e symmetries severely constrains the dynamics, while demanding the nonrelativistic limit \cite{frc5} to be recovered removes any further ambiguity in the multiscale formulation. The case with only Lorentz (and neither translation nor parametrization) invariance is considerably more complicated and we develop it in full only in the uncharged case. 

The coordinates $x^\mu$, however, act also as parameters of fields, which transform under the Poincar\'e algebra of generators according to the rules of tensor calculus. These rules are modified in multiscale spacetimes, and the Poincar\'e algebra is deformed \cite{frc6,frc8}. Thus, when relativistic particles are embedded as sources in field theory, we expect the equations of motion to reflect the multiscale nature of the problem. This is indeed the case, as we will illustrate for a charged particle in Maxwell theory. In the presence of a source $J^\mu(x)$, the energy-momentum tensor of a $U(1)$ gauge field $A_\mu$ is not conserved. To get a conserved tensor density, we must specify the dynamics of the source term. In ordinary Maxwell theory, one adds to the action the contribution of a relativistic charged particle \cite{LL2,Zwi09}. Building on the multiscale version of Maxwell theory \cite{frc8}, here we will do the same, eventually obtaining the expected continuity equation of the full energy-momentum tensor, but only in geometries which are multiscale only in spatial directions. This anisotropic configuration is precisely the one which can effectively describe both noncommutative geometries with cyclicity-preserving measures \cite{ACOS} and Ho\v{r}ava--Lifshitz spacetimes \cite{fra7} as multifractional systems.

Before starting, we comment on one particular multiscale theory where the discussion is simple enough to be summarized in one paragraph. In a model [dubbed ``$q$ theory'' from the use of composite coordinates $q(x)$ here called $\vr$] \cite{frc1,frc2,frc7} invariant under the nonlinear Poincar\'e-like transformation $\vr^\mu({x'}^\mu)=\Lambda^\mu_{\  \nu}\,\vr^\nu(x^\nu)+a^\mu$, all coordinates in the \emph{usual} action are replaced by the distributions
\be\label{xvr}
x^\mu\to \vr^\mu(x^\mu)\,.
\ee
The resulting theory is not a trivial reformulation of ordinary physics because the ``coordinates'' $\vr^\mu$ are actually composite objects and have scale-dependent, multiply anomalous scaling. This means that the physical momentum is canonically associated with the nonanomalous coordinates $x$, not with $\vr$, and the structure of momentum space is different. In this geometry, the line element is
\be\label{svr}
\rmd s_\vr:=\sqrt{-\rmd \vr^\mu(x)\,\rmd \vr_\mu(x)}\,,
\ee
the action for the relativistic neutral particle is simply $S_{\rm p}=-m\int\rmd s_\vr$, and the nonrelativistic action $S_{\rm nonrel}\sim (m/2)\int\rmd\vr^0(t)\,{\vr^i}'{\vr_i}'$ is recovered. Here, roman indices $i=1,\dots,D-1$ run over spatial directions, primes denote derivatives with respect to $\vr^0$, the sum convention for repeated indices is adopted, and we take the Minkowski metric with signature $\eta={\rm diag}(-,+,\cdots,+)$. The equations of motion $m\p_s^2 \vr^\mu(x)=0$ follow suit. If the particle is charged, its coupling with a Maxwell field is straightforward and is formally identical to the standard case under the replacement \Eq{xvr}. Although the $q$ theory can have a number of physical applications \cite{frc1,frc7,fra7}, regarding the specific problem of the relativistic particle we will focus on another multiscale model, namely, the one with weighted Laplacian \cite{frc6,frc7,frc8}, which is mathematically more interesting.

In Sec.\ \ref{unch}, we propose two versions of the uncharged system. The choice between the cases will be determined by various factors, including whether we impose nonrelativistic mechanics as the limit of the more fundamental action $S_{\rm p}$. This will select the anisotropic case, but the two cases will actually collapse into each other when asking compatibility with Maxwell theory (Secs.\ \ref{ch} and \ref{emt}). This restriction (optional, as argued in Sec.\ \ref{disc}) is $v_0(t)=1$, eventually leading back to the Poincar\'e-invariant scenario. Still, the geometry is nontrivial, since it is multiscale along spatial directions.


\section{Uncharged particle}\label{unch}

Consider first a noncharged relativistic particle. In ordinary spacetime, the only Poincar\'e-invariant action recovering the nonrelativistic Lagrangian $L=(m/2)\dot x_i \dot x^i$ is, in $c=1$ units, 
\be\label{Sp1}
\bar S_{\rm p}[x]=-m\int\rmd \bar s\,,\qquad \rmd \bar s[x]:=\sqrt{-\rmd x_\mu\rmd x^\mu}\,,
\ee
where $\rmd \bar s$ is the infinitesimal line element. In other words, the action measures the proper-time interval between the initial and final point of the particle worldline. In time-space components, the line element can be also written as $\rmd \bar s=\rmd t/\bar\g$, where $\bar\g:=1/\sqrt{1-\dot x_i\dot x^i}$ and dots denote derivatives with respect to time $t=x^0$. 

In the context of field theory, the free field action of multiscale spacetimes with weighted Laplacians is invariant under the Poincar\'e algebra of transformations defined by fractional momentum and Lorentz operators \cite{frc6} associated with ordinary translations and Lorentz transformations
\be\label{lore}
{x'}^\mu=\Lambda^\mu_{\  \nu}\, x^\nu
\ee
on the coordinates $x^\mu$. A relativistic particle's worldline is represented by the coordinates themselves; so, when it is added to a field action as a matter content, a natural possibility is to require the particle action $S_{\rm p}$ to be invariant under Eq.\ \Eq{lore}. This considerably limits the form of the dynamics. If we ask for Poincar\'e invariance, we end up with the action \Eq{Sp1}, which, however, does not lead asymptotically to the action of a fractional nonrelativistic particle \cite{frc5}
\be\label{nore}
S_{\rm nonrel}= \frac12 m\sum_i\int \rmd t\, v_0(t)\, ({}_v\cD_t x^i)^2\,,
\ee
where $v_0(t)$ is the measure weight along the time direction and
\be\label{vD}
{}_v\cD_t=\frac{1}{\sqrt{v_0(t)}}\p_t \left[\sqrt{v_0(t)}\,\cdot\,\right]\,.
\ee
This weighted derivative actually stems from the spacetime derivative with the full measure weight $v(x)$,
\be\label{wed}
\cD_\mu:=\frac{1}{\sqrt{v(x)}}\p_\mu \left[\sqrt{v(x)}\,\cdot\,\right]\,.
\ee
Since $v$ is assumed to be factorizable, on a function $f(t)$ one has $\cD_\mu f={}_v\cD_t f$. This derivative is the natural choice in multiscale quantum mechanics giving rise to a self-adjoint momentum operator \cite{frc5}. In order to recover \Eq{nore} for $v_0(t)\neq 1$, we have to give up translation and Lorentz invariance. This is still compatible with field theory, since the action $S_{\rm p}$ is, like Eq.\ \Eq{Sp1}, nonquadratic in the fundamental variables.


\subsection{Isotropic action weight}

\subsubsection{Line element and action}

Multiscale actions feature a Lebesgue measure with a nontrivial weight, and one may ask what happens if one applies the same criterion to augment the action \Eq{Sp1} by a generic positive semidefinite weight function $\tilde\om(s)$:
\be\label{Sp2}
S_{\rm p}=-m\int\rmd s\,\tilde \om(s)\,,\qquad \rmd s:=\sqrt{-\om(s)\,\rmd_\om x_\mu\rmd_\om x^\mu}\,,
\ee
where
\be
\rmd_\om:=\frac{1}{\sqrt{\om(s)}}\,\rmd \left[\sqrt{\om(s)}\,\cdot\,\right]
\ee
is the weighted fractional differential with respect to another weight function $\om(s)$. For the time being, we regard $\om$ and $\tilde \om$ as independent. Notice that $s$ is not the Lorentz distance in multiscale Minkowski spacetime; the actual distance is $s_\vr$ in all versions of the theory, Eq.\ \Eq{svr} \cite{frc1,frc2}. 
The above definition of $\rmd s$ is implicit; after some algebraic manipulation, one finds
\ba
\rmd s&=&\frac{2}{4/\om+\Omega^2\, x^\nu x_\nu}\left\{-\Omega\,x_\mu\rmd x^\mu\vphantom{\sqrt{\frac12}}\right.\nonumber\\
&&\left.+\sqrt{\Omega^2\,(x_\mu\rmd x^\mu)^2-\left[\frac{4}{\om}+\Omega^2\,x^\nu x_\nu\right] \rmd x_\mu \rmd x^\mu}\right\},\nonumber\\\label{tras}
\ea
where $\Omega(s):=\p_s \om/\om$, and the sign in front of the square root has been chosen to get the correct signature in the limit $\Om\to 0$. This expression is transcendental in $s$ and, typically, cannot be inverted to get $s=s(x)$; we will use it once later for a check.

Let us parametrize the worldline $x^\mu(\t)$ of the particle with an arbitrary parameter $\t$, as in the ordinary case \cite{Zwi09}. The action \Eq{Sp2} becomes
\be\label{Sp3}
S_{\rm p}=\int\rmd \t\, w(\t)\,L=-m\int\rmd \t\,\tilde w(\t)\,\sqrt{-w(\t)\,\hat u_\mu\hat u^\mu}\,,
\ee
where the Lagrangian density $L$ has been defined according to the weight carried by the derivatives \cite{frc5}, the functions $\tilde w(\t):=\tilde \om[s(\t)]$ and $w(\t):=\om[s(\t)]$ are fixed profiles of the parameter $s$, and $\hat u^\mu = \cD_\t x^\mu$, with the understanding that $\cD_\t$ has weight $w(\t)=\om[s(\t)]$ and $\cD_s$ has weight $\om(s)$. Therefore, under a reparametrization $\t\to \t'$ they will change form as $w(\t)\to w'(\t')=\om[s(\t')]$, and the action is parametrization invariant only if $\om=1=w$. Here we called $\hat u^\mu$ the $D$-dimensional ``fractional velocity'' vector, using a hat to avoid confusion with the relativistic velocity symbol
\be\label{uu}
u^\mu := \cD_s x^\mu\,,\qquad \om\,u^2=\om\,u_\mu u^\mu \stackrel{\text{\tiny \Eq{Sp2}}}{=}-1\,.
\ee

The measure weight $\tilde w$ can be naively fixed by requiring that the action \Eq{Sp3} in the integer picture reduces to the usual one \Eq{Sp1}. This is not mandatory but it is somewhat expected from the behaviour of the nonrelativistic particle mechanics \cite{frc5} and of field theories \cite{frc6}. In particular, the nonrelativistic case suggests to define \cite{frc5}
\be\label{chiw}
\chi^\mu(\t):=\sqrt{w(\t)}\,x^\mu(\t)\,,
\ee
so that $\rmd s=\sqrt{-\rmd\chi_\mu\rmd\chi^\mu}=\rmd\bar s[\chi]$ and
\be\label{Sp4}
S_{\rm p}[x]=-m\int\rmd \t\,\tilde w\,\sqrt{-\p_\t\chi_\mu\p_\t\chi^\mu}=\bar S_{\rm p}[\chi]
\ee
if, and only if,
\be\label{ww}
\tilde w(\t)=1= \tilde \om(s)\,,
\ee
which will be also required by the dynamics for self-consistency.

\subsubsection{Equations of motion}

We do \emph{not} yet impose Eq.\ \Eq{ww}. Since $\tilde w$ and $w$ are fixed functions of the parameter $\t$, their variation with respect to $\de x^\mu$ when calculating the equations of motion is zero. In fact, from
\ba\nonumber
\de (\rmd s)^2 &=& 2\rmd s\,\de(\rmd s)=-2w \de (\cD_\t x^\mu) \cD_\t x_\mu\, (\rmd\t)^2\\
&=&-2w \cD_\t \de x^\mu \cD_\t x_\mu\, (\rmd\t)^2,\nonumber
\ea
one gets $\de (\rmd s)=-\rmd\t\,w\,u_\mu \cD_\t \de x^\mu$. Applying this equation and the variational principle to the action \Eq{Sp3} with measure \Eq{ww}, after integrating by parts we obtain
\be\label{deS}
\de S_{\rm p} =-\int\rmd\t\,w\,(m\cD_\t u_\mu)\, \de x^\mu
\ee
and the equations of motion
\be\label{cdp}
\cD_\t p_\mu=0\,,\qquad p_\mu:=\tilde \om\,m u_\mu\,.
\ee
The vector $p^\mu$ is the canonical momentum of the relativistic particle, as one can check when treating the latter as a constrained Hamiltonian system \cite{HRT}. Given the Lagrangian density $L=-m(\tilde w/w)\sqrt{-w\hat u^2}$ in \Eq{Sp3},
\ba
p^\mu &:=&\frac{\p L}{\p \hat u_\mu}= \frac{\tilde w}{\sqrt{w}}\frac{m\hat u^\mu}{\sqrt{-\hat u^2}}
= \frac{\tilde w}{\sqrt{w}}\frac{m u^\mu}{\sqrt{-u^2}}\nonumber\\
&\ \stackrel{\text{\tiny \Eq{uu}}}{=}\ &\tilde w\,m u^\mu\,.\label{pcan}
\ea
From the expression $p^2=\tilde\om^2\, m^2 u^2$, one gets the momentum-mass relation
\be\label{1st}
p_\mu p^\mu+m_w^2= 0\,,\qquad m_w=\tilde w\,\frac{m}{\sqrt{w}}=\tilde\om\,\frac{m}{\sqrt{\om}}\,.
\ee
This is not the usual dispersion relation one would have expected from the findings in nonrelativistic mechanics \cite{frc5} or in scalar field theory \cite{frc6}. The ultimate reason is the dynamical role of Eq.\ \Eq{1st}, where $-p^2$ is (the square of) a dynamical canonical variable and not the eigenvalue of the Laplace--Beltrami operator. The presence of an effective varying mass is remindful of the analogous varying electric charge found in Maxwell theory and electrodynamics \cite{frc8}, although the density currents associated with, respectively, the actual mass $m$ and the electric charge will obey crucially different continuity equations.

Let
\be
\check{\cD}_\t:=\frac{1}{w(\t)}\p_\t \left[w(\t)\,\cdot\,\right]
\ee
be the weighted derivative acting on bilinear densities: $\check{\cD}_\t (AB)=B \cD_\t A+A\cD_\t B$. Applying $\check{\cD}_\t$ to Eq.\ \Eq{1st} and using the equations of motion \Eq{cdp}, we uniquely fix the function $\tilde w$ as in Eq.\ \Eq{ww}. Thus, $p_\mu=m u_\mu$, and setting $\t=s$ ($w=\om$) in the equations of motion \Eq{cdp} yields
\be\label{pmu}
\cD_s p^\mu= m\cD_s^2 x^\mu=0\,,
\ee
in agreement with the equations of motion of a massless free particle in nonrelativistic mechanics ($\cD_t^2 x^i=0$) \cite{frc5} and with the Klein--Gordon equation in massless scalar field theory ($\cD_\mu\cD^\mu \phi=0$) \cite{frc6}.

Notice that the integer picture \Eq{Sp4} stemming from the change of variables \Eq{chiw} is consistent. We have seen that the action \Eq{Sp4} corresponds to the one of a standard particle with worldline $\chi^\mu$. Since $\bar s=s$, the momentum associated with $\chi$ is $\bar p^\mu=m\p_{\bar s}\chi^\mu=\sqrt{\om}\,p^\mu$, and Eq.\ \Eq{pmu} coincides with the standard equation $\p_{\bar s}\bar p^\mu=m\p_{\bar s}^2\chi^\mu=0$.

In Hamiltonian formalism, Eq.\ \Eq{1st} is a first-class constraint, which we implement by replacing the symbol $=$ with a weak equality $\approx$ on the constraint surface. As for the relativistic particle in ordinary spacetime, the presence of a constraint stems from the fact that the momentum $p$ and the Lagrangian $L$ are, respectively, zeroth and first order in the ``velocities'' $\hat u$. As a consequence, there is no unique solution $\hat u^\mu(x,p)$, the Lagrangian is singular, and the canonical Hamiltonian $H:= p_\mu \hat u^\mu-L=0$ vanishes identically. The Dirac Hamiltonian can then be written as the sum of the canonical Hamiltonian plus the first-class constraint multiplied times a function $f(\t)$: $H_{\rm D}:= H+f(\t)\,(p_\mu p^\mu+m_w^2)\approx 0$. The fractional Hamilton equations \cite{frc5} correctly stem from this object. The Poisson bracket between momentum and $H_{\rm D}$ yields the equations of motion \Eq{cdp},
\be
\cD_\t p^\mu=\{p^\mu,H_{\rm D}\}=-\frac{\p H_{\rm D}}{\p x_\mu}=-\frac{\p f}{\p x_\mu}(p^2+m_w^2)\approx 0\,,
\ee
while the bracket for $x^\mu$ fixes the function $f$: $\hat u^\mu=\cD_\t x^\mu=\{x^\mu,H_{\rm D}\}=\p H_{\rm D}/\p p_\mu=2fp^\mu$, hence
\be
H_{\rm D}= \frac{w}{2m}\sqrt{-\hat u^2}\,(p^2+m_w^2)\,.
\ee
When $w(\t)=1$, the system is parametrization invariant and the residual arbitrariness in the parameter $\t$ can be removed by considering a gauge constraint $x^0-\t\approx 0$ and treating Eq.\ \Eq{1st} as a second-class constraint \cite{HRT}, but we will not do it here.

\subsubsection{Nonrelativistic limit}

The representation in the integer picture can be used as the starting point to discuss the nonrelativistic limit. Expanding Eq.\ \Eq{Sp4} ($\tilde w=1$) in an inertial frame where $|\p_\t\chi^i/\p_\t\chi^0|\ll 1$, we get
\ba
S_{\rm p} &=&-m\int\rmd \t\,\sqrt{-\p_\t\chi_\mu\p_\t\chi^\mu}\nonumber\\
&=& -m\int\rmd \t\,\sqrt{(\p_\t\chi^0)^2-(\p_\t\chi^i)^2}\nonumber\\
&\approx& -m\int\rmd \chi^0+S_{\rm nr}\,,\label{24}
\ea
where we assumed, without loss of generality, that $\p_\t \chi^0>0$ and
\ba
S_{\rm nr} &=& \frac12 m\sum_i\int \rmd \t\, \frac{1}{\p_\t \chi^0}\,(\p_\t \chi^i)^2\nonumber\\
&=& \frac12 m\sum_i\int \rmd t\, \frac{1}{\dot\chi^0}\,v_0(t)\,({}_v\cD_t x^i)^2\,,\label{nore2}
\ea
where we set $w[\t(t)]=v_0(t)$ for any identification of the worldline parameter $\t$ with a function of time $x^0=t$. The limit \Eq{24} is also obtained from Eq.\ \Eq{tras}, dropping $O(\Omega^2)$ and $O(x_i^3)$ terms.

To recover the nonrelativistic limit \Eq{nore}, one should impose $\dot\chi^0(t)=\p_t(\sqrt{v_0}\,t)=1$, which implies $w(\t)=1=v_0(t)$ and a spacetime whose multiscale structure is encoded only in spatial directions. This result motivates a different scenario, where the measure weight $w\to w_\mu$ in the derivatives is anisotropic.


\subsection{Anisotropic action weight}

\subsubsection{Line element and action}

A generalization of the previous case is obtained by allowing for the differential $\rmd_\om$ in Eq.\ \Eq{Sp2} to carry different weights $\om_\mu$ for each of the $D$ directions:
\be
(\rmd_\om x)^\mu:=\frac{1}{\sqrt{\om_\mu(s)}}\,\rmd \left[\sqrt{\om_\mu(s)}\,x^\mu\right]\,.
\ee
The notation of the previous section can be used, provided expressions of the form $w \hat u^2$ and similar are replaced by their counterpart where the sum over spacetime indices is extended also to the weights, $w \hat u^2\to w\cdot \hat u\cdot \hat u=\sum_\mu w_\mu \hat u^\mu \hat u_\mu$. For instance, the action can be written as ($\tilde \om=1$ from the start)
\ba
S_{\rm p}&=&-m\int\rmd s=-m\int\sqrt{-\rmd(\sqrt{\om_\mu}\,x_\mu)\,\rmd(\sqrt{\om_\mu}\,x^\mu)}\nonumber\\
&=&-m\int\rmd \t\,\sqrt{w\cdot\hat u\cdot \hat u}\,,\qquad \hat u^\mu=(\cD_\t x)^\mu\,.\label{Sp5}
\ea
Consequently, the normalization \Eq{uu} of the velocity holds, $ \om\cdot u\cdot u=-1$. We call $\om_\mu$ the ``action weights'' to avoid confusion with the spacetime weight $v(x)$. A system with anisotropic action weights is not necessarily associated with a spacetime with anisotropic spacetime weight.

\subsubsection{Equations of motion}

The variation $\de (\rmd s)=-\rmd\t\,w\cdot u\cdot\cD_\t \de x$ 
yields the equations of motion
\be\label{cdp2}
(\cD_\t p)^\mu=0\,,\qquad p^\mu= m (\cD_s x)^\mu\,.
\ee
In particular, for $\t=s$ one has
\be\label{pmu2}
(\cD_s p)^\mu= m(\cD_s^2 x)^\mu=0\,.
\ee
The momentum $p$ can be also defined by the generalization of Eq.\ \Eq{pcan} working in the integer picture and computing $\bar p^\mu=\sqrt{\om_\mu}\,p^\mu$.

The dispersion relation \Eq{1st} cannot be cast in terms of $p^2$, but, rather, with a weighed squared momentum:
\be\label{1st2}
\om\cdot p\cdot p+m^2= 0\,.
\ee


\subsubsection{Nonrelativistic limit}

Reconciling Eqs.\ \Eq{nore} and \Eq{nore2} is now straightforward. It is sufficient to make the following anisotropic identifications of the action weights, such that derivatives acting on $x^0$ are normal and those acting on $x^i$ all have the same weight:
\be
\om_0=1=w_0\,,\qquad \om_i[s(t)]=v_0(t)=w_i[\t(t)]\,.
\ee
Then, $\dot\chi^0(t)=\p_t(\sqrt{w_0}\,t)=1$ and $S_{\rm nr}= S_{\rm nonrel}$. The line element of this system is
\bs\label{es}\be
\rmd s=\sqrt{\rmd t^2-v_0(t)\,(\rmd_v x^i)^2}=\frac{\rmd t}{\g}\,,
\ee
where
\be
\g=\frac{1}{\sqrt{1-v_0(t)\,({}_v\cD_t x^i)^2}}\,.
\ee\es
The nature of the nonrelativistic approximation $\g\sim 1$ in multiscale spacetimes depends not only on how much the modulus of the velocity $\dot {\bf x}$ is smaller than the speed of light $c=1$, but also on the time when the approximation is taken with respect to the evolution of the Universe determined by the hierarchy of scales contained in the measure weight (and which can be better appreciated in the multifractional realization of these measures \cite{frc6,frc7,frc4,frc8}).

The case of anisotropic geometries with standard time direction is of particular interest. Setting $v_0=1$ makes the relativistic-particle system with line element \Eq{es} ordinary ($\chi^\mu=x^\mu$ for all $\mu$) and Poincar\'e invariant, Eq.\ \Eq{Sp1}. The nontrivial multiscale structure along spatial directions [$v_i(x^i)\neq 1$] can be seen only in quantum mechanics \cite{frc5} or when fields are coupled to the particle, as in the case of electromagnetism that we will illustrate below. In fact, the application of ordinary Lorentz transformations to the coordinates labeling fields will be instrumental to couple the relativistic particle and multiscale field theory consistently.

Ultimately, the complications of the case with isotropic action weights stem from the fact that, for the relativistic particle, the time coordinate is also a degree of freedom of the system. In contrast, in nonrelativistic mechanics time is a parameter and spatial coordinates are the degrees of freedom, while in field theory all coordinates are parameters and the degrees of freedom are tensorial densities. In other words, the coordinate or field redefinitions mapping the fractional picture to the integer picture [e.g., $\chi^i(t)=\sqrt{v_0(t)}\, x^i(t)$ in nonrelativistic mechanics and $\cA_\mu=\sqrt{v(x)}\,A_\mu$ in electrodynamics] do not combine different degrees of freedom, since the measure weights are functions of parameters. On the other hand, in relativistic mechanics the relation \Eq{chiw} does entangle different degrees of freedom when moving to a parametrization choice $\t=f(t)$ or $\t=f(s)$.

\section{Charged particle}\label{ch}

When the relativistic particle interacts with fields, there is no clean representation in the integer picture except in anisotropic geometries with $v_0(t)=1$. We can see this in electromagnetism. The Maxwell action in multiscale spacetimes is \cite{frc8}
\be\label{clf}
S_F=-\frac1{4}\int\rmd^Dx\,v(x)\,F_{\mu\nu} F^{\mu\nu}\,,
\ee
where
\be
F_{\mu\nu}   =\cD_\mu A_\nu-\cD_\nu A_\mu\label{maxf}
\ee
is the field strength of the Abelian gauge field density vector $A$. To recast the action in the integer picture, it is sufficient to make the field redefinition $A_\mu\to \cA_\mu=\sqrt{v(x)}\,A_\mu$, so that $S_F=(-1/4)\int\rmd^Dx\,\cF_{\mu\nu} \cF^{\mu\nu}$, where $\cF_{\mu\nu} =\p_\mu \cA_\nu-\p_\nu \cA_\mu=\sqrt{v(x)} F_{\mu\nu}$. However, at the level of particle mechanics the integer picture entails a change of coordinates, Eq.\ \Eq{chiw}, which should not happen in a field-theory context. Combined with the above field redefinition, the net result is considerably more intricate. First, we notice that (sum over three-time repeated index $\s$)
\be
\frac{\rmd}{\rmd x^\mu}=\frac{\rmd\chi^\s}{\rmd x^\mu}\frac{\rmd}{\rmd \chi^\s}=\sqrt{w_\s}\frac{\cD_\t x^\s}{\p_\t x^\mu}\frac{\rmd}{\rmd \chi^\s}=:\xi_{\mu\s}(x,\t)\frac{\rmd}{\rmd \chi^\s}\,,
\ee
so that
\ba
F_{\mu\nu} &=& \frac{1}{\sqrt{v(x)}}\left(\p_\mu \cA_\nu-\p_\nu \cA_\mu\right)\nonumber\\
					 &=& \frac{1}{\sqrt{v(x)}}\left[\xi_\mu^{\, ~\s}(x,\t)\frac{\p\cA_\nu}{\p\chi^\s}-\xi_\nu^{\, ~\s}(x,\t)\frac{\p\cA_\mu}{\p\chi^\s}\right].
\ea
The expression in brackets coincides with $\cF_{\mu\nu}$ only if $\xi_\mu^{\,~\s}=\de_\mu^\s$. This happens only when $w_\mu=1$ for all $\mu$. In particular, the case where $w_0=1$ and $w_i=v_0(t)$ is compatible with the integer-picture field theory if, and only if,
\be\label{vw}
v_0(t)=1\,,\qquad w_\mu(\t)=1\,,
\ee
consistently with the nonrelativistic limit. This corresponds to a geometry which is multiscale \emph{only along spatial directions}. The alternative to imposing this condition is to forfeit the integer-picture field theory altogether. This is not an issue \emph{per se}, since the physics is defined in the fundamental variables $x^\mu$ and field $A_\mu$, while the integer picture is only a useful calculational tool which, as known, fails in interacting systems \cite{frc6}. Here we have the double complication of having both an interaction (between the particle and the electromagnetic field) and a system (the relativistic particle) which is nonlinear even when taken alone, at least at the level of the action. Overall, it would seem reasonable to abandon the integer picture. 

However, one may raise at least three objections against doing so. First, the electromagnetic interaction should at most produce an effective spacetime-dependent electric charge while allowing for the integer picture \cite{frc8}. Second, the equations of motion of the relativistic particle are indeed linear in the coordinates. Last, failing to impose Eq.\ \Eq{vw} renders the equations of motion and the treatment of the energy-momentum tensor far more complicated. Therefore, from now on we do assume \Eq{vw}.

We will keep the same velocity symbols as before, with the understanding that $\hat u^\mu=\p_\t x^\mu$ and $u^\mu=\p_s x^\mu$. A charged particle produces an electromagnetic field $A_\mu[x^\nu(\t)]$ according to the action $S_{\rm p}+S_e$, where
\be\label{Se}
S_e=\int\rmd x^\mu\, \tilde e A_\mu=\int\rmd \t\,\hat u^\mu\, \tilde e A_\mu \,.
\ee
Here, $\tilde e=e_0\sqrt{v(x)}$ is the effective spacetime-dependent charge appearing in the covariant derivatives of electrodynamics \cite{frc8} and $e_0$ is the electron charge.
The equations of motion can be easily found following almost the usual steps \cite{LL2,Zwi09}. Varying the action \Eq{Se} with respect to $\de x^\mu$,
\ba
\de S_e &=& \int\rmd\t\left[\hat u^\nu\de(\tilde e A_\nu)+\tilde e A_\mu\de(\p_\t x^\mu)\right]\nonumber\\
				&=& \int\rmd\t\left[\hat u^\nu\p_\mu (\tilde e A_\nu)\de x^\mu+\tilde e A_\mu\p_\t \de x^\mu\right]\nonumber\\
        &=& \int\rmd\t\left[\hat u^\nu\p_\mu (\tilde e A_\nu)\de x^\mu-\p_\t(\tilde e A_\mu) \de x^\mu\right]\nonumber\\
				&=& \int\rmd\t\,\hat u^\nu\left[\tilde e\,(\cD_\mu A_\nu)\de x^\mu-\p_\nu(\tilde e A_\mu) \de x^\mu\right]\nonumber\\
				&=& \int\rmd\t\,\hat u^\nu\,\tilde e F_{\mu\nu}\,\de x^\mu\,.
\ea
Adding this to Eq.\ \Eq{deS} with $w=1$, one obtains the equations of motion
\be\label{deS2}
\p_\t p_\mu=\hat u^\nu\tilde e F_{\mu\nu}\,,
\ee
or, for $\t=s$, $m\p_s u_\mu= u^\nu\tilde e F_{\mu\nu}$.

Equation \Eq{Se} can be written in terms of a source $J^\mu_e$. First, one replaces $\tilde e$ with a spatial charge distribution $\rho_e$ of pointwise charges $\tilde e_n\propto \tilde e({\bf x}_n)$ located at various points ${\bf x}_n$. In our case, $\rho_e$ should take into account the nontrivial measure weight of space, so that the ordinary Dirac distribution is replaced by a multiscale spatial delta distribution $\de_v({\bf x},{\bf x}_n):=\prod_i\de(x^i-x^i_n)/\sqrt{v_i(x^i)\,v_i(x^i_n)}$ (e.g., Ref.\ \cite{frc4}). Multiplying $\rho_e$ times $\dot x$ yields the charge density current
\be
J^\mu_e:=\rho_e\,\dot x^\mu\,,\qquad \rho_e=\sum_n\tilde e_n\de_v({\bf x},{\bf x}_n)\,.
\ee
Thus, $\tilde e\to \rmd\tilde e:=\rho_e\,\rmd\vr({\bf x})$ in Eq.\ \Eq{Se}, which becomes ($\t=t$)
\be\label{Se2}
S_e=\int\rmd x^\mu\, \rmd\tilde e A_\mu=\int\rmd^Dx\,v(x)\, J^\mu_e A_\mu\,.
\ee
In particular, in the integer picture one has $S_e=\bar S_e=\int\rmd^Dx\, \cJ^\mu_e \cA_\mu$, where $\cJ^\mu_e:=\sqrt{v}\,J^\mu_e$. The equations of motion \Eq{deS2} can be recast in terms of mass and charge densities,
\be\label{deS3}
\rho_m\p_s u_\mu=\rho_e u^\nu F_{\mu\nu}\,,
\ee
where $\rho_m=\sum_n m_n\de_v({\bf x},{\bf x}_n)$.


\section{Energy-momentum tensor}\label{emt}

To get the energy-momentum tensor \cite{LL2}, one notices that the four-momentum density should be of the form ${}^{({\rm p})}T^{0\nu}=\rho_m u^\nu$, so that
\be
p^\nu =\int\rmd\vr({\bf x})\,{}^{({\rm p})}T^{0\nu}\,.
\ee
Moreover, the mass density is the $0$ component of $\rho_m \dot x^\mu$, so that the symmetric energy-momentum tensor of the relativistic particle is
\be\label{Tp}
{}^{({\rm p})}T^{\mu\nu}:=\rho_m \g u^\mu u^\nu=J_m^\mu  u^\nu\,,
\ee
where $\g=\rmd s/\rmd t$ and in the second step we defined a mass density current $J_m^\mu:=\rho_m\dot x^\mu$. Despite its similarity with the charge density current, they obey different continuity laws. Let
\be
\check{\cD}_\mu:=\frac1v \p_\mu \left[v\,\cdot\,\right]
\ee
be the weighted spacetime derivative for bilinears. In the present case, $\check{\cD}_0=\p_t$. While it was shown in Ref.\ \cite{frc8} that the charge density current is conserved with respect to the weighted derivative \Eq{wed}, $\cD_\mu J_e^\mu=0$, due to conservation of the total mass $M=\sum_n m_n$ we have
\be\label{JJ}
0=\check{\cD}_\mu J_m^\mu= \dot \rho_m+ \frac{1}{v_i}\,\p_i (v_i\,J_m^i)\,.
\ee
Integrating over the spatial volume and throwing away a boundary term, one gets $\dot M=0$.

The multiscale continuity equation for ${}^{({\rm p})}T$ is obtained by applying these results:
\ba
\check{\cD}_\mu\, {}^{({\rm p})}T^\mu_{\, ~\nu} &=& u_\nu\check{\cD}_\mu J_m^\mu+ J_m^\mu \p_\mu u_\nu\ \stackrel{\text{\tiny \Eq{JJ}}}{=}\ J_m^\mu \p_\mu u_\nu\nonumber\\
																					 &=& \rho_m \dot x^\mu \frac{\rmd s}{\rmd x^\mu} \p_s u_\nu\ \stackrel{\text{\tiny \Eq{deS3}}}{=}\ -\rho_e \dot x^\mu \frac{\rmd s}{\rmd x^\mu} u^\mu F_{\mu\nu}\nonumber\\
																					 &=& -J_e^\mu F_{\mu\nu}\label{Tpcon}\,.
\ea
If we add the dynamics for the $U(1)$ gauge field $A$, we obtain conservation of the energy-momentum tensor of Maxwell theory in multiscale spacetimes with pointwise particle sources. The total action is $S=S_{\rm p}+S_e+S_F$. The energy-momentum tensor associated with the Maxwell term is \cite{frc8}
\be\label{emt2}
{}^{(F)}T_{\mu\nu}=-\frac{1}{4}F^{\s\tau} F_{\s\tau}\eta_{\mu\nu} +F_\mu^{\ \s}F_{\nu\s}\,,
\ee
which, upon using the cyclic relation $\check{\cD}_\s F_{\mu\nu}+\check{\cD}_\mu F_{\nu\s}+\check{\cD}_\nu F_{\s\mu}=0$ and Maxwell equations $\cD_\nu F^{\mu\nu}=J^\mu_e$, obeys the (non)conservation law \cite{frc8}
\be\label{TFcon}
\check{\cD}_\mu\, {}^{(F)}T^\mu_{\, ~\nu}=J^\mu_e F_{\mu\nu}\,.
\ee
Combining this with Eq.\ \Eq{Tpcon}, we get
\be
\check{\cD}_\mu\,T^\mu_{\, ~\nu}=\check{\cD}_\mu\, [{}^{(F)}T^\mu_{\, ~\nu}+{}^{({\rm p})}T^\mu_{\, ~\nu}]=0\,,
\ee
as announced in \cite{frc8}.


\section{Discussion}\label{disc}

We conclude by drawing some consequences for the spectral dimension $\ds$ of multiscale spacetimes \cite{frc7} and electrodynamics \cite{frc8}.

The spectral dimension of a geometry is obtained by letting a test particle diffuse via a transport equation which can be constructed from stochastic nonrelativistic mechanics. While in ordinary geometries this procedure is straightforward, in multiscale spacetimes there is an ambiguity in the choice of the scaling of the abstract diffusion time. In turn, this scaling determines the value of the spectral dimension \cite{frc7}. If the scaling is nonanomalous, $\ds=D$. In this paper, we found that the charged relativistic particle admits an integer picture and is compatible with field theory only in spatially multiscale geometries where time direction is ordinary. An ordinary nonrelativistic limit, as implied by Eq.\ \Eq{vw}, would tell us that the diffusion equation should be standard and, thus, $\ds=D$.

The same reasoning would also lead to the conclusion that uniform charge distributions in multiscale spacetimes with $v_0(t)=1$ are associated with a measured constant electric charge. This would rule out the effects of the time-varying fine-structure constant of multiscale origin discussed in Ref.\ \cite{frc8}. Yet, nontrivial effects coming from a space-dependent electric charge (hence, a spatially varying fine-structure constant) would not be excluded.

All these conclusions are based upon the nonrelativistic limit \Eq{nore} and compatibility of the charged relativistic particle with field theory. Concerning the first assumption, a simple alternative avoiding the condition $\ds=D$ is to give up the nonrelativistic limit \Eq{nore}, assume the relativistic formulation as fundamental, and define the nonrelativistic limit through it, as in Eq.\ \Eq{nore2}. Then, one would not need to impose constraints on the action weights, but the nonrelativistic action would nevertheless lead to a modified mechanics and a modified diffusion equation with respect to that found in Ref.\ \cite{frc7}. On the other hand, if one insists to keep Eq.\ \Eq{nore}, one should also remark that the particle-plus-fields system is not fundamental, in contrast with a pure field-theory approach where the electromagnetic source producing the current $J_e$ is a fermionic density field \cite{frc8}. At a level more fundamental than classical mechanics, the requirement \Eq{vw} is not compelling. This suggests to turn around the above conclusions and reinterpret them in a more conservative way: Electromagnetic sources may be well approximated by the relativistic mechanics presented here only in multiscale geometries with ordinary time direction. When time is anomalous, one should resort either to field theory or to a more complicated mechanics model.


\bigskip

\begin{acknowledgments}
The author thanks Giuseppe Nardelli and especially David Rodr\'iguez for useful discussions. This work is under a Ram\'on y Cajal tenure-track contract.
\end{acknowledgments}


\end{document}